\input harvmac
\input epsf

\newcount\figno
\figno=0
\def\fig#1#2#3{
\par\begingroup\parindent=0pt\leftskip=1cm\rightskip=1cm\parindent=0pt
\baselineskip=12pt
\global\advance\figno by 1
\midinsert
\epsfxsize=#3
\centerline{\epsfbox{#2}}
\vskip 14pt

{\bf Fig. \the\figno:} #1\par
\endinsert\endgroup\par
}
\def\figlabel#1{\xdef#1{\the\figno}}
\def\encadremath#1{\vbox{\hrule\hbox{\vrule\kern8pt\vbox{\kern8pt
\hbox{$\displaystyle #1$}\kern8pt}
\kern8pt\vrule}\hrule}}

\overfullrule=0pt

\noblackbox
\parskip=1.5mm

\def\Title#1#2{\rightline{#1}\ifx\answ\bigans\nopagenumbers\pageno0
\else\pageno1\vskip.5in\fi \centerline{\titlefont #2}\vskip .3in}

\font\caps=cmcsc10

\noblackbox
\parskip=1.5mm

  
\def\npb#1#2#3{{ Nucl. Phys.} {\bf B#1} (#2) #3 }
\def\plb#1#2#3{{ Phys. Lett.} {\bf B#1} (#2) #3 }
\def\prd#1#2#3{{ Phys. Rev. } {\bf D#1} (#2) #3 }
\def\prl#1#2#3{{ Phys. Rev. Lett.} {\bf #1} (#2) #3 }
\def\mpla#1#2#3{{ Mod. Phys. Lett.} {\bf A#1} (#2) #3 }
\def\ijmpa#1#2#3{{ Int. J. Mod. Phys.} {\bf A#1} (#2) #3 }

\def\cmp#1#2#3{{ Commun. Math. Phys.} {\bf #1} (#2) #3 }

\def\bb#1{{[arXiv:hep-th/#1]}}

\def\heph#1{{[arXiv:hep-ph/#1]}}

\def\atmp#1#2#3{{ Adv. Theor. Math. Phys.} {\bf #1} (#2) #3 }
\def\jhep#1#2#3{{ J. High Energy Phys.} {\bf #1} (#2) #3 }


\def\CL{{\cal L}}   
   
\def\CM{{\cal M}} \def\CP{{\cal P}}  
\def\CN{{\cal N}}


\def\dj{\hbox{d\kern-0.347em \vrule width 0.3em height 1.252ex depth
-1.21ex \kern 0.051em}}

\def\Tr{{\rm Tr\,}}

\def\Dirac{\,\raise.15ex\hbox{/}\mkern-13.5mu D}
\def\dirac{\,\raise.15ex\hbox{/}\kern-.57em \partial}
\def\aslash{\,\raise.15ex\hbox{/}\mkern-13.5mu A}

\def\shalf{{\ifinner {\textstyle {1 \over 2}}\else {1 \over 2} \fi}} 
\def\sshalf{{\ifinner {\scriptstyle {1 \over 2}}\else {1 \over 2} \fi}} 
\def\sfourth{{\ifinner {\textstyle {1 \over 4}}\else {1 \over 4} \fi}}

\lref\rlus{M. Luscher, \npb{219}{1983}{233.}}

\lref\rtho{G. 't Hooft, \npb{138}{1978}{1.} \npb{153}{1979}{141.} \npb{205}{
1982}{1.}} 

\lref\rpotl{M. Luscher and G. Munster, \npb{232}{1984}{445.} J. Koller
and P. van Baal, \npb{273}{1986}{387.} \prl{57}{1986}{2783.} \npb{302}{1988}{1.}
P. van Baal, \npb{307}{1988}{274,} [Erratum-ibid {\bf B312} (1989) 752].}

\lref\rindex{E. Witten, \npb{202}{1982}{253.} \atmp{5}{2002}{841} \bb{0006010}.} 

\lref\rrev{P. van Baal, Acta Phys. Polon. {\bf B20} (1989) 295. \heph{0008206}.
A. Gonz\'alez Arroyo, \bb{9807108}.}  

\lref\rvbi{P. van Baal, \bb{0112072}.}

\lref\rgpy{D.J. Gross, R.D. Pisarsky and L.G. Yaffe, Rev. Mod. Phys. {\bf 53}
(1981) 43.}
 
\lref\rorb{S. Kachru an E. Silverstein, \prl{80}{1988}{4855} \bb{9802183}.
 A. E. Lawrence, N. Nekrasov and C. Vafa, \npb{533}{1998}{199}
\bb{9803015}.
M. Bershadsky, Z. Kakushadze and C. Vafa, \npb{523}{1998}{59}
\bb{9803076}. Z Kakushadze, \npb{529}{1998}{157} \bb{9803214}. \prd{58}{1998}{
106003} \bb{9804184}.
  M. Bershadsky and A. Johansen, \npb{536}{1998}{141} \bb{9803249}.
 M. Schmalz, \prd{59}{1999}{105018} \bb{9805218}. } 

\lref\rstras{M. J. Strassler, \bb{0104032}.}

\lref\rori{A. Armoni, M. Shifman and G. Veneziano, \npb{667}{2003}{170} 
\bb{0302163}. \bb{0403071}. Fortsch. Phys. {\bf 52} (2004) 453. F. Sannino
and M. Shifman, \prd{69}{2004}{125004} \bb{0309252}. A. Armoni, A. Gorsky
and M. Shifman, \npb{702}{2004}{37} \bb{0404247.} P. di Vecchia, A. Liccardo,
R. Marotta and F. Pezzella, \bb{0412234}. F.~Sannino, \bb{0507251}.}  

\lref\radic{A. Armoni, M. Shifman and G. Veneziano, \prl{91}{2003}{191601}
\bb{0307097}. \plb{579}{2004}{384}
\bb{0309013.}}

\lref\radip{A. Armoni, M. Shifman and G. Veneziano, \prd{71}{2005}{045015}  
\bb{0412203}.} 
 
\lref\rcesar{E. Cohen and C. G\'omez, \npb{223}{1983}{183}. 
\plb{135}{1984}{99}. \prl{52}{1984}{237}.}  

\lref\rcont{A. Gorsky and M. Shifman, \prd{67}{2003}{022003} \bb{0208073}.
R. Dijkgraaf, A. Neitzke and C. Vafa, \bb{0211194}. M. Erlich and A. Navqi,
\jhep{0112}{2002}{047} \bb{9808026}. P. Kovtun, M. Unsal and L.G. Yaffe,
\jhep{0312}{2003}{034} \bb{0311098}. \bb{0411177}. \bb{0505075}.  
A. Armoni, A. Gorsky and M. Shifman, \bb{0505022}.} 

\lref\runsal{ P. Kovtun, M. Unsal and L.G. Yaffe, \bb{0505075}.}

\lref\rtong{D. Tong, \jhep{0303}{2003}{022} \bb{0212235}.}

\lref\rdinkin{S. Okubo and J. Patera, J. Math. Phys. {\bf 25} (1984) 219.
T. van Ritbergen, A. N. Schellekens and J.A.M. Vermaseren, \ijmpa{14}{1999}{41}
\heph{9802376}.}
 
\lref\rdin{K.R. Dienes, \npb{429}{1994}{533}  
\bb{9402006}.}  

\lref\rcarlosp{C. Hoyos, in preparation.}

\lref\rlargen{G. 't Hooft, \npb{75}{1974}{461.}}

\lref\rmaldacobi{N. Itzhaki, J. Maldacena, J. Sonnenschein and S. Yankielowicz,
\prd{58}{1998}{046004} \bb{9802042.}
}

\lref\rme{J.L.F. Barb\'on,  \plb{543}{2002}{283}  
\bb{0206207.}}  

\lref\rads{E. Witten,
{\it Adv. Theor. Math. Phys.} {\bf 2} (1998)
253 \bb{9802150.} S.S. Gubser, I.R. Klebanov and A.M. Polyakov,
\plb{428}{1998}{105} \bb{9802109.}}

\lref\rexpri{J. Maldacena and A. Strominger, \jhep{9812}{1998}{005}
\bb{9804085.}}

\lref\rHP{G.T. Horowitz and J. Polchinski, \prd{55}{1997}{6189}
\bb{9612146.}}

\lref\rberkooz{M. Berkooz, A. Sever and A. Shomer, \bb{0112164.}}

\lref\rwitm{E. Witten, \bb{0112258.}}

\lref\rotros{P. Minces and V.O. Rivelles, \jhep{0112}{2001}{010} \bb{0110189.}
W. Muck,
\plb{531}{2002}{301}
\bb{0201100.}
P. Minces,
\bb{0201172.}
A.C. Petkou,
\bb{0201258.}
E.T. Akhmedov,
\bb{0202055.}
A. Sever and  A. Shomer,
\bb{0203168.}}

\lref\rwitmast{E. Witten, {\it ``Recent Developments in Gauge Theories"}, 1979
Cargese Lectures. Ed. G. 't Hooft et. al. Plenum (1980).}

\lref\rexodef{O. Aharony, M. Berkooz and E. Silverstein, \jhep{0108}{2001}{006}
\bb{0105309.} \bb{0112178.}}

\lref\rthresholds{J.L.F. Barb\'on, I.I. Kogan and E. Rabinovici,
\npb{544}{1999}{104} \bb{9809033.}}

\lref\rHPage{S.W. Hawking and D. Page, \cmp{87}{1983}{577.}}

\lref\rwitheta{E. Witten, \prl{81}{1998}{2862}
\bb{9807109.}}

\lref\rwitthp{E. Witten, {\it Adv. Theor. Math. Phys.} {\bf 2} (1998)
505 \bb{9803131.}}

\lref\rmalda{J. Maldacena, {\it Adv. Theor. Math. Phys.} {\bf 2} (1998)
231 \bb{9711200.}}

\lref\roldmm{S.R. Das, A. Dhar, A.M. Sengupta and S.R. Wadia, \mpla{5}{1990}{1041.}
L. Alvarez-Gaum\'e, J.L.F. Barb\'on and C. Crnkovic, \npb{394}{1993}{383.} G.
Korchemsky, \mpla{7}{1992}{3081,} \plb{256}{1992}{323.} I.R. Klebanov,
\prd{51}{1995}{1836.} I.R. Klebanov and A. Hashimoto, \npb{434}{1995}{264.}
J.L.F. Barb\'on, K. Demeterfi, I.R. Klebanov and C. Schmidhuber, \npb{440}{1995}{189.}}


\baselineskip=15pt

\line{\hfill IFT UAM/CSIC-2005-31}
\line{\hfill FTUAM-2005-10}
\line{\hfill {\tt hep-th/0507267}}

\vskip 1.0cm

\Title{\vbox{\baselineskip 12pt\hbox{}
 }}
{\vbox {\centerline{Small volume expansion of almost    }
\vskip10pt
\centerline{supersymmetric large $N$ theories}
}}

\vskip1.0cm

\centerline{$\quad$ {\caps 
J.L.F. Barb\'on
 and   
C. Hoyos 
}}
\vskip0.7cm

\centerline{{\sl  Instituto de F\'{\i}sica Te\'orica C-XVI,  
 UAM, Cantoblanco 28049. Madrid, Spain }}
\centerline{{\tt jose.barbon@uam.es, c.hoyos@uam.es}}

\vskip0.2cm


\vskip1.0cm

\centerline{\bf ABSTRACT}

 \vskip 0.3cm

 \noindent 

We consider the small-volume dynamics
of nonsupersymmetric orbifold and orientifold
field theories defined on a three-torus, in a test 
of  the claimed planar equivalence
between these models and appropriate supersymmetric ``parent models".
We study  one-loop 
effective potentials over the moduli space of flat connections
and  find that planar equivalence is preserved for suitable averages over the
moduli space. On the other hand, strong nonlinear effects produce
local violations of planar equivalence 
at special points of moduli space. 
 In the case of orbifold models, these effects show that the
 ``twisted" sector dominates
the low-energy dynamics.

\vskip 1.5cm

\Date{July 2005} 
               

\vfill





\baselineskip=15pt

\newsec{Introduction}

\noindent

Nonsupersymmetric gauge theories with an effectively supersymmetric large-$N$
limit have been the subject of considerable recent interest. The main
examples involve the so-called orbifold \refs\rorb\ and orientifold \refs\rori\ 
field
theories. The prototype of the first class of models is a $U(N)^k$ quiver
theory with bi-fundamental fermions and ${\bf Z}_k$ global symmetry, whose restriction  
to planar diagrams is equivalent, up to coupling rescalings, to the analogous
planar-diagram approximation of a $U(kN)$ Super Yang--Mills (SYM) 
model with minimal $\CN =1$ supersymmetry. 
 In the case of orientifold field theories,  $SU(N)$ gauge theory  
with Dirac fermions in  two-index representations (either symmetric or antisymmetric) is
claimed to be planar-equivalent to the corresponding $\CN =1$, $SU(N)$  gauge theory. In
both cases the statement of planar equivalence must be restricted to a particular
class of observables that can be appropriately mapped from ``parent" to ``daughter" theories
under the orbifold/orientifold projection. A similar correspondence must be established
among the vacua of the theories in question, a very important point given the potentially
complicated vacuum structure of these models.

A stronger version  of this perturbative correspondence is the hypothesis of nonperturbative
planar equivalence (c.f. \refs\rstras), namely the parent and daughter theories would
have equivalent sectors to leading order in the large-$N$ expansion but non-perturbatively
in the 't Hooft coupling, $\lambda = g^2 N$. If true, this stronger version of the 
planar equivalence would yield interesting predictions for nonperturbative quantities 
in non-supersymmetric gauge theories, by a modified $1/N$ expansion whose first term is
protected by supersymmetry. A remarkable example  of this program 
is the calculation of the chiral condensate of one-flavour nonsupersymmetric 
QCD (c.f. \refs\radic), in terms of the gaugino condensate of the $\CN=1$  SYM parent
gauge theory. 

There have been a number of arguments and
counter-arguments regarding this important question, especially in the context of
the orbifold field theories \refs\rcont.
 In particular, a crucial necessary condition for the
strong version of planar equivalence to hold is that the global symmetry group ${\bf Z}_k$
should not be spontaneusly broken in either the parent or the  
daughter vacua under consideration. 
This restriction leaves $k=2$ as the only candidate orbifold model to realize the
strong version of the planar equivalence conjecture (c.f. \refs\runsal), although the calculation of
\refs\rtong\ shows that, under compactification on ${\bf R}^3 \times {\bf S}^1$, 
the effective three-dimensional theory breaks spontaneously the crucial ${\bf Z}_2$ symmetry.  
On the other hand, the situation is much better for the case of orientifold planar
equivalence, where master-field arguments seem to provide a formal proof of the correspondence
(c.f. \refs\radip).  

In this paper we study the phenomenon of planar equivalence in a small-volume expansion
on a three-torus (see \refs\rrev\ for a review).
 We take spacetime to be ${\bf T}^3 \times {\bf R}$ with the ${\bf R}$
factor representing time, and  ${\bf T}^3$ a straight torus of size $L$. 
  In the limit of a small torus, there is
a clean Wilsonian separation of scales between ``slow" degrees of freedom, i.e. zero modes on
${\bf T}^3$, and the non-zero modes, i.e. the ``fast" variables in the language of
the Born--Oppenheimer approximation.    
For asymptotically free theories, this limit is accessible to weak-coupling methods,
and a systematic effective action over the configuration space of the zero modes can
be constructed (c.f. \refs{\rlus, \rpotl}).
 Our purpose is to use  these well-developed techniques to get
some insight on the question of non-perturbative planar equivalence. 

The paper is organized as follows. In section 2 we consider the constraints induced
by the hypothesis of planar equivalence on the graded partition function, the natural
generalization of the Witten index for these theories. In section 3 we review the
details of the Born--Oppenheimer approximation for gauge theories on tori. In section 4
we apply these results to the basic examples of orbifold/orientifold models and use them to  
test the claim of planar equivalence. Finally, we offer some concluding remarks
in section 5. 

\newsec{Planar equivalence and the ``planar index"}  

\noindent

The naturally protected quantity for minimally supersymmetric systems in finite
volume is the supersymmetric
 index, $I= \Tr (-1)^F$, \refs\rindex. In the case at hand, we shall be interested
in the related graded partition function  
\eqn\indt{
I(\beta) = \Tr \; \CP \; (-1)^F \; e^{- \beta \,H}\;,}
where $\CP$ is a projector introducing possible  refinements of the  index with
respect to extra global symmetries of the problem.  In principle, a judicious choice
of $\CP$ might be necessary to 
 establish the planar equivalence, but we shall suppress it for the time being.  
Unlike the analogous object in the
parent $\CN=1$ theory, $I(\beta)$ is not independent of $\beta/L$ or the dimensionless
couplings in the Lagrangian. The interesting question  is
whether we can establish an approximate BPS character of $I(\beta)$.

The strongest possible statement of planar equivalence would have 
$I(\beta)$ behaving
as a supersymmetric index of a rank $N$ supersymmetric gauge theory,
at least in some dynamical limit. This would mean a large-$N$
scaling of order
\eqn\indp{
I(\beta) = O(N)
\;.
}  
On the other hand,
  perturbative  planar equivalence poses much weaker constraints on the graded
partition function, i.e. it only requires   
\eqn\st{ \log\; I(\beta) = O(N) \;,}
since the leading $O(N^2)$ term is a sum of planar diagrams and 
should vanish as in the  supersymmetric parent.  
 Notice that  condition 
\st\ is exponentially weaker than \indp.
The physical interpretation of the minimal planar equivalence condition
\st\ depends to a large extent on the dynamical 
regime that we consider. For example, the implications for the
structure of the spectrum   depend on the value of the 
ratio $\beta/L$. 
Let us  write 
\indt\ in terms of the spectrum of energy eigenvalues, 
\eqn\prj{
I(\beta) = \sum_{E} \;\left( \Omega_B (E) - \Omega_F (E)\right)\; e^{-\beta
\,E}\,,}
with $\Omega_{B,F}$ the  boson and fermion density of states.
The thermal partition function  is given by
\eqn\tprj{
Z \,(\beta) = \sum_{E} \;\left( \Omega_B (E) + \Omega_F (E)\right)\; e^{-\beta
\,E}\,.}
In the large-volume limit, $\beta /L \ll 1$, these quantities are dominated
by the high-energy asymptotics of the spectrum. In particular, for 
asymptotically free theories, we expect that the free energy can be 
approximated by that of a plasma phase. Thus, on dimensional   
 grounds, 
$$
\log\,Z \,(\beta) = -N^2 \,f (\lambda) \,(L/\beta\,)^3 + O(N) 
\;,
$$
up to a vacuum-energy contribution linear in $\beta$, with $f(\lambda)$ a function
of the 't Hooft coupling $\lambda=g^2 N $, conveniently renormalized at the scale
$1/\beta$.  From this expression we infer that
the asymptotic high-energy behaviour of the density of states is given by
$$
\log\,\Omega_{B,F} (E) = \sqrt{N} \,s_{B,F} (\lambda) \,(E\,L)^{3/4} + O(N) 
\,.$$
Notice that the leading term is of $O(N^2)$ for $E = O(N^2)$. Then, the
prediction of planar equivalence boils down to $s_B = s_F$, namely the leading
$O(N^2)$ high-energy asymptotics of the density of states is effectively supersymmetric. \foot{
This property is very similar to the so-called ``misaligned supersymmetry", studied in
\refs\rdin, that characterizes non-supersymmetric string theories without classical  
tachyonic instabilities.} 

Conversely, in the opposite limit $L/\beta \ll 1$ the graded partition function $I(\beta)$
is dominated by low-lying states.
In the classical approximation, the Hamiltonian of gauge theories on ${\bf T}^3$ has
 states of vanishing potential energy,
 corresponding to flat connections on the gauge sector and zero eigenvalues
of the Dirac operator on the fermion sector. This implies that there is a neat
separation of scales, of order $1/L$,   between zero modes and the rest of the degrees of freedom in the
limit of small volume. The graded partition function may be written as
\eqn\gpz{
I(\beta) = \Tr_{\rm slow}  \, (-1)^F\, e^{-\beta H_{\rm eff}} \,,
}
with $H_{\rm eff}$ a Wilsonian  effective Hamiltonian for the zero modes or ``slow variables".
For asymptotically free theories, this Hamiltonian
may be estimated in a weak-coupling expansion
in the small running coupling $g^2 N$, defined at the scale
$1/L$ (c.f. \refs{\rlus, \rpotl}).

A characteristic behaviour of the supersymmetric index
is that the effective Hamiltonian acting on the space of zero modes with energies much
smaller than $1/L$ is free, i.e. the Wilsonian  effective potential induced by integrating out
the non-zero modes vanishes exactly as the result of boson/fermion  cancellation.
In the non-supersymmetry  case, a similar behaviour of \gpz\ would require that the
effective potential be at most 
of order $V_{\rm eff} = O(1/N)$, i.e. it would have to
 vanish in the large $N$ limit. On the other
hand, planar equivalence would  require the weaker condition that   
this effective potential vanishes to leading $O(N^2)$ order or, equivalently $V_{\rm eff} =
O(N)$. 

Therefore, the existence of a sort of ``planar index"  seems to be a stronger property
than ``minimal" planar equivalence, in the sense discussed in the Introduction. 
Understanding this fact in detail is important, given the close relationship between
the supersymmetric index in finite volume 
and the gaugino condensates in the standard $\CN=1$ SYM lore 
\refs{\rcesar}.

 In this paper we examine the zero-mode effective potentials for particular examples
 of theories in which a form of planar equivalence is expected to hold. To be more precise,
 we consider  the basic ${\bf Z}_2$ orbifold model, 
 with gauge group $SU(N)\times SU(N)$ and a Dirac bi-fundamental fermion. 
  We also 
 consider the orientifold  model, an $SU(N)$ gauge theory with a Dirac fermion
 in either the  symmetric or antisymmetric two-index representations.

\newsec{Effective potentials for flat connections}

\noindent

We focus on gauge theories with simply-connected gauge group on the torus,
namely, we can take periodic boundary conditions for all the fields involved,  generating 
the maximal possible  set of zero modes. 
The configuration space of bosonic zero modes is the moduli space of flat connections
on ${\bf T}^3$, which  in turn can be characterized by a commuting triple of holonomies, modulo
gauge transformations (c.f. \refs\rrev). 
If the gauge group is taken to be a direct product of $SU(N)$ factors,
the moduli space has the same product structure, with each factor given by direct product
of Cartan tori ${\bf T}_{\rm C}^{N-1}$ 
for each holonomy, modulo the Weyl group $W=S_N$, which acts by permutations of
the holonomies'   eigenvalue spectrum,\foot{In what follows, we focus on 
gauge groups with simple  $SU(N)$ factors, neglecting for example
 the $U(1)$ factors that arise
in  orbifold models, which would only contribute subleading effects in
the large $N$ limit.}   
$$
\CM_{SU(N)}= {\left({\bf T_{\rm C}}^{N-1} \right)^3  
 \over W}
\;.$$
As coordinates on $\CM$, we choose  constant dimensionless
gauge fields in the Cartan
subalgebra $L\,{\vec A} =
\sum_{a=1}^{N-1} {\vec C}^a \,H^a$. We refer to the  torus
$({\bf T_{\rm C}}^{N-1})^3$ as the ``toron valley",  defined as  
the product of three copies of  ${\bf R}^{N-1}/2\pi {\widetilde \Lambda}_r $, where
${\widetilde \Lambda}_r$ is the dual of the root lattice, i.e. we identify ${\vec C}$  modulo   
translations by   $4\pi \alpha\, {\bf Z}^3$, with $\alpha$ a root and  ${\bf Z}^3$ a 
three-vector
of integers. 
 The Weyl group acts by reflections on the hyperplanes orthogonal to the
roots, and makes $\CM$ into an orbifold. 

Fermionic zero modes are given by solutions of $\Dirac \psi =0$.  At a generic point on
$\CM$, flatness of the gauge field implies that $\psi$ is actually 
  covariantly constant, i.e. 
 invariant under the  holonomies that parametrize $\CM$. If
$\psi$ is in the adjoint representation of the gauge group, this puts the fermion zero modes
on the Cartan subalgebra for generic points on $\CM$. However, for either fundamental or
symmetric/antisymmetric representations, fermion zero modes will be supported on submanifolds
of $\CM$ of zero measure. For this reason, when discussing the non-supersymmetric theories,
we shall consider only the bosonic zero modes
as explicit variables and integrate out all the fermionic degrees of freedom. 

Under these circumstances, the effective Lagrangian around a  {\it generic} point of 
the bosonic moduli space
takes the form
\eqn\eflag{
\CL_{\rm eff} =  {L\over 2g^2} \,\Tr \, \left(\partial_t\, {\vec C}\,\right)^{\,2} -
V_{\rm eff} (\,{\vec C}
\,)
\,,}
where the effective potential is the sum of bosonic and fermionic parts. In the one-loop approximation 
and in background-field gauge it is given by
\eqn\polt{
\int_{-\infty}^\infty dt\;
 V_{\rm eff} ({\vec C}\,) = 
\Tr_{\rm Ad} \;\log\,\left(-D^2_{\vec C}\,\right)-\sum_R \,\Tr_R \;\log\,\left(\,i\,\Dirac_{\vec C}
\,\right)\;,}
where $R$ stands for the fermion representations. The operator
 $-D^2$ acts on scalar functions on ${\bf R}\times {\bf T}^3$,
 after we have taken into
account the contribution from gauge-field polarizations and ghosts. 
The normalization convention for $i\Dirac$  regards fermion representations 
as carried by
Majorana or Weyl spinors, so that representations carried by Dirac spinors 
actually
involve $R\oplus {\bar R}$, contributing with an extra  factor of $2$  
to the total potential.

Using the fact that $(i\Dirac)^2 = -D^2$ for flat connections, we simply need to compute
the determinant of the operator $-D^2$ in different representations. Since we work on
the space of constant abelian connections, the traces in \polt\ involve a sum over the 
relevant weight spaces and we can write    
(c.f. \refs{\rlus, \rpotl, \rrev}) 
\eqn\potl
{V_{\rm eff} ({\vec C}\,) = \sum_{\alpha } V\left(\alpha \cdot  
{\vec C} \right) - \sum_R \sum_{\mu \in R } V\left( \mu\cdot  {\vec C} 
\right)\;,}
where $\alpha$ runs over the roots of $SU(N)$, $\mu$ runs over the weights of the
fermion representations  $R$ and the dot product refers to the 
Cartan inner product, i.e. $\alpha \cdot {\vec C} \equiv \sum_{a=1}^{N-1} \,\alpha_a {\vec C}_a$.
In this general expression, the function $V$ is the subtracted
 scalar determinant 
$$
\int_{-\infty}^\infty dt\; V({\vec \xi}\;) = \Tr \;\log\; \left[-\partial_t^2 -
\left({\vec \partial} + L^{-1} \,{\vec \xi}\,\right)^2 
\right]_{{\bf R} \times {\bf T}^3} - 
\Tr \;\log\; \left[-\partial^2 \,\right]_{{\bf R} \times {\bf T}^3}  
\;,
$$
which can 
 be computed by standard methods (c.f. \refs{\rlus, \rgpy})
  with the result  
\eqn\podos{
V({\vec \xi} \;) = {2\over \pi^2 L} \,
\sum_{{\vec n} \neq 0} 
{\sin^2 \left(\shalf \,{\vec n} \cdot {\vec \xi}\, \right) 
\over \left({\vec n}\cdot {\vec n}\,\right)^2 } \;,}
up to numerical,  ${\vec \xi}$-independent constants. We shall refer to
\podos\ as the $SU(2)$ potential, since it coincides with the contribution,   
 up to a factor of $2$, of  the $SU(2)$ adjoint representation.

These expressions make it clear that the parent
 supersymmetric theories have vanishing effective
potential over $\CM$, since fermions are in the adjoint representation and the bosonic
and fermionic potentials cancel out one another. 

\subsec{General properties of the effective potential}

\noindent

The elementary building block of \potl, the $SU(2)$ potential, 
has periodicity ${\vec \xi} \rightarrow
{\vec \xi} + 2\pi {\vec k}, \;{\vec k} \in {\bf Z}^3$ and is symmetric under reflections
${\vec \xi} \rightarrow -{\vec \xi}$. The minima of the positive function 
$V({\vec \xi}\;)$ are
located at ${\vec \xi}_{\rm min} = 0$ modulo $2\pi {\bf Z}^3$, and correspond to conical
singularities where $V({\vec \xi}\;) \sim  |{\vec \xi}-{\vec \xi}_{\rm min} |$. 
Maxima
are smooth and sit at 
${\vec \xi}_{\rm max} = (\pi,\pi,\pi) $ modulo $2\pi {\bf Z}^3$ (c.f. Fig 1). 

\fig{\sl  $SU(2)$ bosonic potential along a Cartan direction. Conical minima are
located at $C=0\;{\rm mod}\; 2\pi$, 
smooth maxima at $C=\pi\;{\rm mod}\;2\pi$.
 }{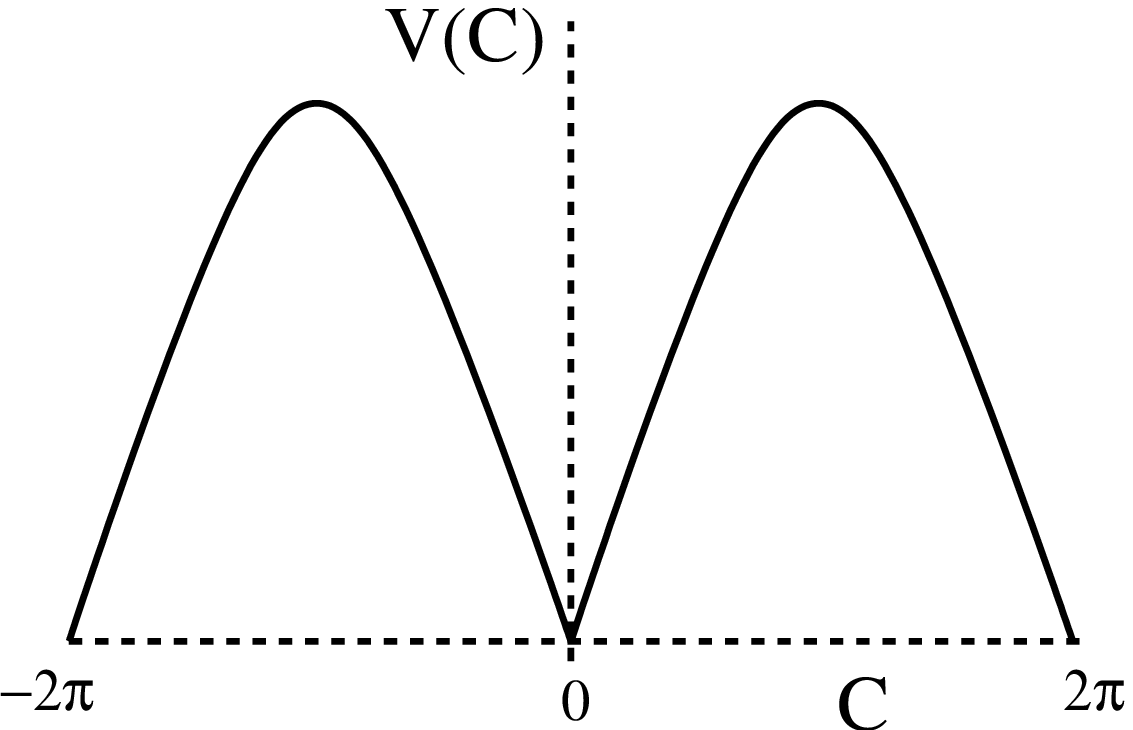}{2truein}

These properties propagate to the  full $SU(N)$ potential, with due consideration of the
global group-theory structure for each contributing representation. 
The potential induced by  the pure-gauge degrees of freedom
 enjoys a translational symmetry under ${\vec C} \rightarrow {\vec C} + 4\pi {\vec k}\, \nu^i$
with $\nu^i$ any of the $N$ weights of the defining representation. In particular,
all zeros of the adjoint potential can be obtained from ${\vec C}=0$ by the
action of this translational symmetry. The induced  periodicity
on the moduli space is  finer than the minimal one,   imposed by the toroidal
nature of the potential valleys. This periodicity is associated to the action of
the global symmetry group $({\bf Z}_N)^3$ of central conjugations, whose characters label
the non-abelian electric flux sectors \refs\rtho.

In general, this symmetry of central conjugations is broken by the fermionic representations
down to a subgroup of order $n(R)^3$, where $n(R)$ is the so-called ``N-ality" of the
representation. It can be defined by the number of Young tableaux, modulo $N$, or in
the highest-weight parametrization
$$
\mu_R^{\rm hw} = \sum_{k=1}^{N-1} q_k \, \mu^k\;,
$$
with $\mu^k = \sum_{i=1}^k \nu^i$ the fundamental weights and $q_k \geq 0$, the
``N-ality" is given by $n(R) = \sum_k q_k k$ mod $N$. 
Hence, for fermion representations with trivial center, such as the defining ``vector"
representation, the translations by multiples of $4\pi \nu^i$ do not define a symmetry
of the full effective potential.

\fig{\sl  
The structure of the $SU(3)$ potential for 
 the adjoint representation (on the left) and the defining vector  representation
(on the right). The
Weyl hyperplanes are indicated by dashed lines. We see that
the effective Weyl chamber for the vector representation is bigger
 than  the adjoint one 
 by a factor $\sqrt{3}$. 
The conical points (global minima or maxima ) lie on the
intersection of Weyl hyperplanes. For the vector representation, we also
mark by white circles  
the smooth critical points of the potential.
}{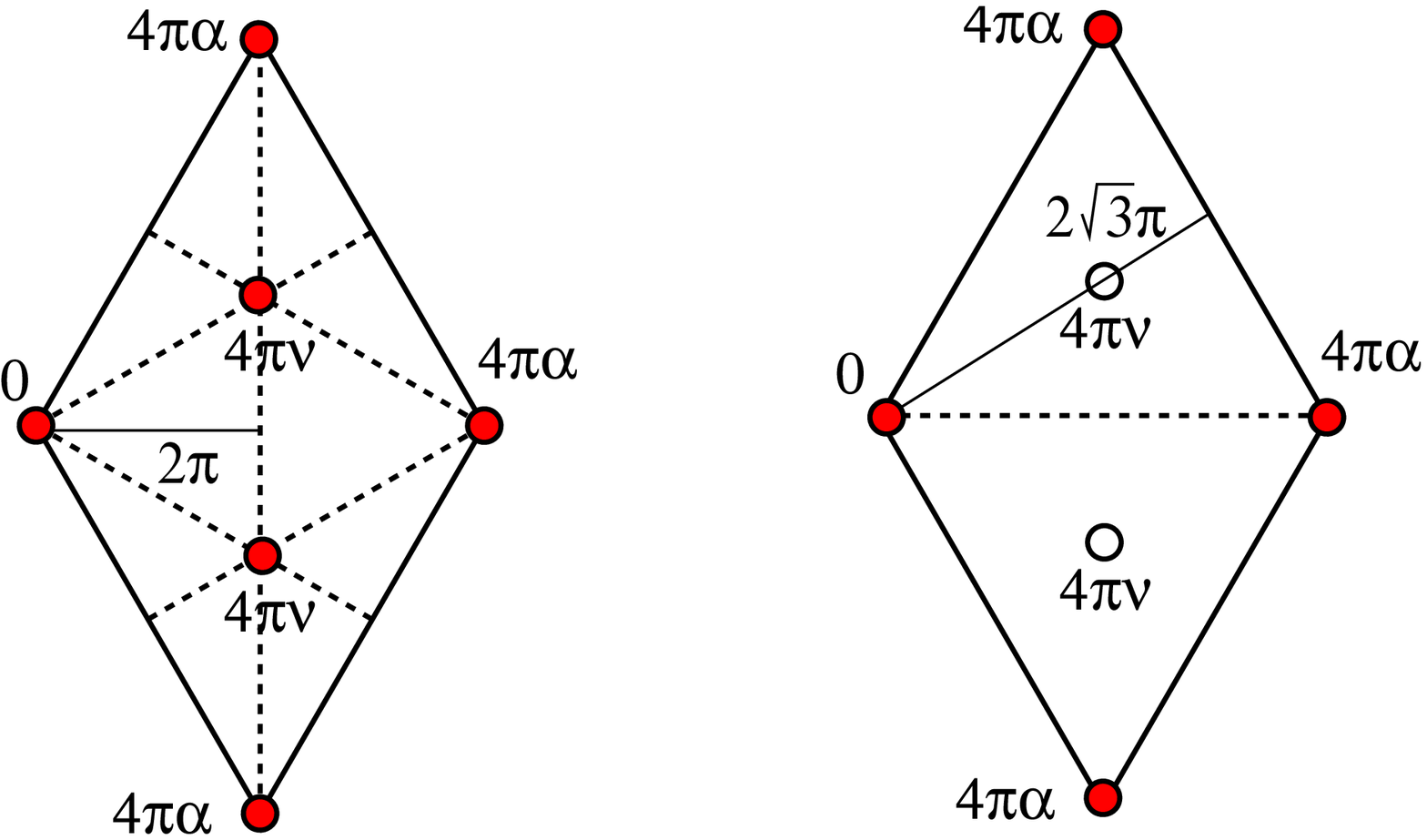}{3truein}

In general, representations with trivial N-ality will induce potentials that fit 
within the frame of the bosonic potential, and partial cancellations of some bosonic 
minima are possible. Representations with non-trivial N-ality will generate potentials
that fit better within the frame of the fundamental (vector) representation. The local
bosonic mimima will generically survive as global minima, except perhaps the one at
the origin.

The contribution of a representation $R$ to the potential is very constrained geometrically
by the action of the Weyl group, generated by permutations of the weights. The combination
of the Weyl reflections and the periodicity properties of the potential single out the
$(N-2)$-dimensional hyperplanes given by the linear equations
$$
\mu \cdot {\vec C} = 0 \;\;{\rm mod}\;\; 2\pi\,{\bf Z}^3
\,.$$
for each weight, there is one such Weyl hyperplane passing through the origin, 
and all others are parallel and shifted by integer multiples of the normal vector $2\pi \mu /
|\mu|^2$. Weyl hyperplanes are local minima of the bosonic potential along the transverse
directions (and local maxima of the fermionic potential). Hence, local minima (maxima) of the
bosonic (fermionic) potential 
are located at intersections between Weyl hyperplanes. The fundamental region of the
Weyl reflection group is the Weyl chamber, and is enclosed by such Weyl hyperplanes. 
As a result local minima (maxima) of the bosonic (fermionic) potential are located at
the vertices of the Weyl chambers. 
For a given representation, the induced potential shows structure up to the scale given by
the size of the associated Weyl chamber,
$$
\ell_R = {2\pi \over |\mu_R |} 
$$
with $\mu_R$ the highest weight of the representation $R$.

 As it stands, the effective potential \potl\ is not smooth at the orbifold singularities
 of $\CM$. The vertices  of the Weyl chamber signal either bosonic minima or fermionic maxima. 
 Locally, the behaviour of the potential at these points is conical
 $V_{\rm eff} \sim |{\vec C}\,|$, a result of  the
 appearance of extra localized light modes. Focusing on the minima, the
 leading non-derivative potential
 away from the toron valley is given by the commutator term
$
\Tr\;[C_\mu, C_\nu]^2/g^2
$, 
which induces a mass for generic excitations orthogonal to the toron valley. Zero-point
fluctuations of these modes are at
the origin of the  effective potential considered above. However,
near ${\vec C}=0$ (or any of the other orbifold singularities), the $N^2 -1$ constant modes
are
effectively massless. Upon rescaling ${\vec C} \rightarrow g^{2/3} {\vec C}$, we obtain
a homogenous
scaling of the effective Hamiltonian yielding a spectrum of order 
 $g^{2/3} /L \sim 1/L N^{1/3}$, with wave functions localized over a region of size
$O(g^{2/3}) \sim O(1/N^{1/3})$ around the orbifold points \refs\rlus.
Away from this ``stadium", the effective potential produces a barrier that 
supresses the wave functions exponentially at large $N$. In general, barriers are very
efficient in supressing tunneling at large $N$, since the flat connections have a large
effective mass  of order $m_{\rm eff} \sim LN/\lambda$ at weak coupling, as
can be seen from \eflag.

These dynamical considerations show that, unless the cancellation between 
bosonic and
fermionic potentials is accurate up to corrections of  $O(1/N)$, 
the low-lying wave functions will tend to
remain rather localized in the large $N$ limit, a situation that is altogether very
different from the supersymmetric case, where zero-mode wave functions can be
considered uniformly distributed over the moduli space (in fact, constant) despite
the existence of orbifold singularities (see \refs{\rvbi, \rindex} for further
discussion of the subtleties involved). 

\newsec{Effective potentials and planar equivalence}

\noindent

At the level of the effective potentials considered in this paper, the property of
planar equivalence manifests itself in the cancellation of bosonic and fermionic
contributions
  to leading order in the large-$N$ expansion. Namely, one should find  
$V_{\rm eff} ({\vec C}\,) = O(N)$ instead of the more generic $O(N^2)$ scaling.  Alternatively,
one relates the given model to its supersymmetric parent. In many cases, this
comparison is rather subtle. For example,  
 in orbifold models the detailed
 global structure of the toron valleys changes from
partent to daughter, since the Weyl groups are clearly different. 
  Hence, the domains of definition of $V_{\rm eff} ({\vec C}\,)$
 differ between parent and 
  daughter theories and a precise comparison would involve a further refinement of
 the projector that appears in \indt. 

For models with the same configuration space, such as the orientifold
field theory,  the main property that ensures planar equivalence is
the equality, to leading $O(N^2)$ order, of the fermion effective actions between
parent and daughter, i.e.
\eqn\eqpd{
 \;\Tr_{\rm Ad}  \;\log \; (\,i\Dirac\,)-  
\Tr_{\rm R}  \;\log \; (\,i\Dirac\,)    
= O(N)\;.}
In principle, this property may be established in  a strong sense, for any sufficiently
smooth gauge connection $A$, provided it
belongs to the configuration space of both theories.
Within the weak-field expansion, 
the  perturbative version of planar 
equivalence ensures \eqpd\ to any finite order in the one-loop Feynman diagram expansion in
the background field. To see this, consider the one-loop diagram with $n$ external legs,
proportional to
$
\Tr_{\rm R} \left({\aslash / \dirac}\right)^n
$. 
It contains a group-theory factor proportional to $\Tr_{\rm R} \,T^{a_1} \dots T^{a_n}$.
Reducing this trace to a combination of symmetrized traces, we end up with terms of the
form 
$$
{\rm STr}_{\rm R} \;T^{a_1} \dots T^{a_n} = I_n (R) \,d^{\,
a_1 \cdots a_n}+ {\rm lower\,\, order\;\;
products,} 
$$
where 
$I_n (R)$ denotes the Dynkin 
index of the representation $R$, and the
 symmetric polynomial $d^{\,a_1 \cdots a_n}$ is the symmetrized
trace in the fundamental representation.

Then, the Dynkin index for the symmetric and antisymmetric
representations, $S_\pm$,   
 is given by 
 $I_n (S_\pm) = 2(N\pm 2^n)$   
(c.f.
\refs\rdinkin), to be compared with
  $I_n ({\rm Ad}) = 2N$
for the adjoint representation.  
The leading Dynkin index is indeed the same 
in the large-$N$ limit, so that a given diagram with an arbitrary (but {\it fixed}) number
of legs yields  the same contribution in the large-$N$ limit in the parent
and daughter theories. So far
this is just another way of looking at the statement of perturbative planar equivalence.
However,  
$I_n (S_\pm)$ 
has a subleading (non-planar) 
term that blows-up exponentially with the number of legs
of the diagram. Hence, planar equivalence stated diagram by diagram no longer 
guarantees
that the full non-perturbative effective action will satisfy the planar-equivalence
property \eqpd. In principle, there could be a non-commutativity between the large-$N$
limit and the non-linearities beyond the weak-field expansion, and this question could
be a dynamical one, i.e. depending on the particular gauge connection considered as background.

These considerations show that  the behaviour of the exact one-loop
potential \potl\ under planar
equivalence is a rather non-trivial issue. The purely perturbative version 
of the  planar equivalence
is strictly related to the Taylor expansion of the potential around the origin
of moduli space ${\vec C}=0$, and the behaviour at finite distance away
from the origin,  $|{\vec C}| \sim 1$, remains an open question.  In the following subsections we consider
in more detail the effective potential in the two main examples of planar equivalence. 

\subsec{Orientifold effective potential}

\noindent

The orientifold model defined by a $SU(N)$ gauge theory with a Dirac fermion in the
antisymmetric representation has an effective potential for the flat connections  that
we may conveniently write as  a superposition
of the basic $SU(2)$ potentials \podos, following the general structure of \potl.

It is convenient to parametrize the weights and
roots of the relevant representations in terms of the $N$ weights, $\nu^i$, 
of the defining vector
representation of $SU(N)$, that satisfy $\nu^i \cdot \nu^j =  (N\delta^{ij} - 1)/2N$,
so that $\nu^i + \nu^j $ with $i< j$ run over the weights of the antisymmetric
two-index representation, whereas  $\nu^i - \nu^j$ run over
the roots of $SU(N)$ as $i,j = 1, \dots, N$, except for the fact that in this
way we count one extra vanishing root, which gives no contribution since
$V(0)=0$ in our additive convention for the vacuum 
energy. \foot{The 
${\vec C}$-independent part of $V_{\rm eff}$ vanishes to $O(N^2)$ in all the models
studied in this paper.}
The final form of the potential is 
\eqn\orip{
V_{\rm ori} ({\vec C}) = \sum_{i, j} V\left((\nu^i - \nu^j)\cdot {\vec C}\,\right) -
2\,\sum_{i< j} V\left((\nu^i + \nu^j)\cdot {\vec C}\,\right)\;,}
with an extra term
$$
-2 \sum_i V\left(2\nu^i \cdot {\vec C}\,\right)
$$
to be added for the symmetric two-index representation. The first two terms are nominally
of $O(N^2)$ from the multiplicity of the index sums, whereas the last term is at most
of $O(N)$, and thus may be ignored when discussing
 the property of (minimal) 
planar equivalence in this
model. 

The first nontrivial property of the orientifold potential is the
insatbility of the origin of moduli space, ${\vec C}=0$. To see this,
we can explore the potential along the direction of a fundamental
weight, say ${\vec C}=\nu^1{\vec c}$, with very small ${\vec c}$, so
that the $SU(2)$ potential can be approximated by $V({\vec \xi})\simeq
v |{\vec \xi}|$
with $v$ a positive constant.

Evaluating the potential in this direction one obtains
$$
V_{\rm ori} (\nu^1{\vec c}) \approx -N |{\vec c}|
$$
up to subleading corrections at large $N$. Hence, we conclude that the
zero-holonomy point at the origin of moduli space becomes severely
unstable at large $N$.

The scale over which  the bosonic and fermionic contributions vary is
 roughly the same,
since the size of the Weyl chamber for the two-index representations 
is
$$
\ell_{S_\pm} = {2\pi \over |\nu^i + \nu^j|} = 2\pi + O(1/N), \;\;i\neq j\,,
$$
whereas the size of the standard Weyl chamber 
is $\ell_{\rm Ad} = 2\pi / |\nu^i - 
\nu^j |
= 2\pi$.
On the other hand, the alignment of the bosonic and fermionic potentials
is not perfect. 
We can see this by checking the height of the local conical minima~\foot{The conical
singularities are smoothed out when considering the effects of the light non-abelian
degrees of freedom, as explained at the end of Section 3.}, that
are inherited from the absolute zeros of the bosonic potential. 
Conical minima of the bosonic potential are determined by the equations
$$
(\nu^i -\nu^j ) \cdot {\vec C}^{(0)} = 0 \;{
\rm mod}\;2\pi {\bf Z}^3
$$
for all $i,j$, which have as the general solution the integer lattice generated
by the vectors $4\pi \nu^i$.  We can now evaluate the fermionic contribution
$$
V_{\rm F} ({\vec C}^{(0)}) = -2\sum_{i<j} V\left((\nu^i + \nu^j )\cdot {
\vec C}^{(0)}\,\right)
$$
at a generic zero of the bosonic contribution, given by ${\vec C}^{(0)} = 4\pi \sum_i 
{\vec n}_i \,\nu^i$, with ${\vec n}_i \in {\bf Z}^3$. Using the properties of
the basic weights we have  
$$
(\nu^i + \nu^j) \cdot 4\pi\sum_k {\vec n}_k \,\nu^k = 2\pi({
\vec n}_i + {\vec n}_j ) - {4\pi \over N} \sum_k {\vec n}_k
$$
and, by the periodicity properties of the $SU(2)$ potential, we can write  
$$
V_{\rm F} ({\vec C}^{(0)})  \sim  -N^2 V\left({4\pi \over N}
\sum_k {\vec n}_k \right) 
\;.$$
Hence, as we move around the lattice of  conical minima from points with
$\sum_k {\vec n}_k =O(1)$ to points with $\sum_k {\vec n}_k =O(N)$, the
potential scans on a band of  energies ranging from $O(N)$  up to
$O(N^2)$, with a typical spacing of $O(N)$.   As an example of a set of
local minima with $O(N^2)$ negative energy, we can consider points in the
lattice satisfying  $\sum_k {\vec n}_k =(N/4, 0,0)$ at large values of $N$
such that $N/4$ is an integer. On these points,
the fermionic contribution  is maximal and given by 
$$
V_F = - N^2 \,V(\pi) + O(N) 
\;.$$
In an analogous fashion, the conical maxima of the fermionic potential should
be lifted up by the smooth portions of the bosonic potential giving us an 
intrincate {\it landscape} of very narrow valleys and peaks on the $O(N)$ 
scale of relative heights, with  some walls  rising up to
$O(N^2)$ energies. The conclusion is that nonlinear effects at finite distance
away from the origin of moduli space tend to spoil the property of planar
equivalence, at least when looking at the potentials with high enough resolution.

In fact, it turns out that planar equivalence is still maintained on the average, since
the integral of \orip\ over the toron valley is only of $O(N)$. To see this, we can
expand the flat connections in the basis of the simple roots $\alpha_s^i = \nu^i - 
\nu^{i+1}, 
i= 1, \dots, N-1$, i.e. we write
$$
{\vec C} = \sum_{l=1}^{N-1} {\vec c}_l \;\alpha_s^{l}
$$
and the orientifold potential takes the form  
\eqn\oriav{
V_{\rm ori} ({\vec C}\,) = \sum_{i,j} \left[\,V\left(\shalf ({\vec c}_i - {\vec c}_{i-1} -
{\vec c}_j + {\vec c}_{j-1} 
)\right) -  V\left(\shalf ({\vec c}_i -{\vec c}_{i-1} + {\vec c}_j -{\vec c}_{j-1} 
)\right)\right] + O(N)
\;.}
In this expression, we have neglected terms of $O(N)$ coming from  restrictions on the   range of
the indices. When  averaging over the toron valley, the coordinates
${\vec c}_j$ become dummy integration variables, and an appropriate change of variables
shows that the bosonic and fermionic terms cancel out upon integration. 

We can  use similar arguments to show that the average of the squared potential 
$(V_{\rm ori})^2$
 is
at most of $O(N^3)$. This shows that the average cancellation of $O(N^2)$ terms in
\oriav\ does not come from large ``plateaus" of $O(N^2)$ energy occupying different $O(1)$
fractions of moduli-space's volume. If that would have been the case, all these
smooth regions would contribute to $(V_{\rm eff})^2$ as
positive plateaus of $O(N^4)$ energy, producing an average
squared potential of $O(N^4)$. 

The vanishing of  the leading $O(N^4)$ terms in the averaged squared potential 
 suggests that
the valleys and peaks of the orientifold potential are localized over small volumes
of the moduli space in the large $N$ limit. 
Still, the  wave functions will  show strong localization  
properties and our analysis sheds no new light on the important question of
chiral symmetry breaking in these models (non-trival expectation values of
fermion bilinears in the infinite volume limit).

\subsec{Orbifold effective potential}

\noindent

For the basic orbifold model with $SU(N) \times SU(N)$ group and  a bi-fundamental
Dirac fermion we have a potential defined over the direct product of toron valleys
parametrized by ${\vec C}_1$ and ${\vec C}_2$ flat connections. It has the form  
\eqn\orbbp{
V_{\rm orb} ({\vec C}_1, {\vec C}_2) = \sum_{i,j} V\left((\nu^i - \nu^j\,) \cdot {\vec C}_1
\right) +
\sum_{i, j} V\left((\nu^i - \nu^j\,) \cdot {\vec C}_2 \right) - 2\sum_{i, j} V\left(\nu^i \cdot
{\vec C}_1 -
\nu^j \cdot {\vec C}_2\,\right).}

\fig{\sl Section of the potential for the
 $SU(2)\times SU(2)$ orbifold field theory. 
The diagonal line is the region where bosonic and fermionic potentials 
cancel one another. Global
minima of negative energy lie at the points marked by triangles pointing
downwards. Global maxima of positive energy are also indicated by
triangles pointing upwards.
 }{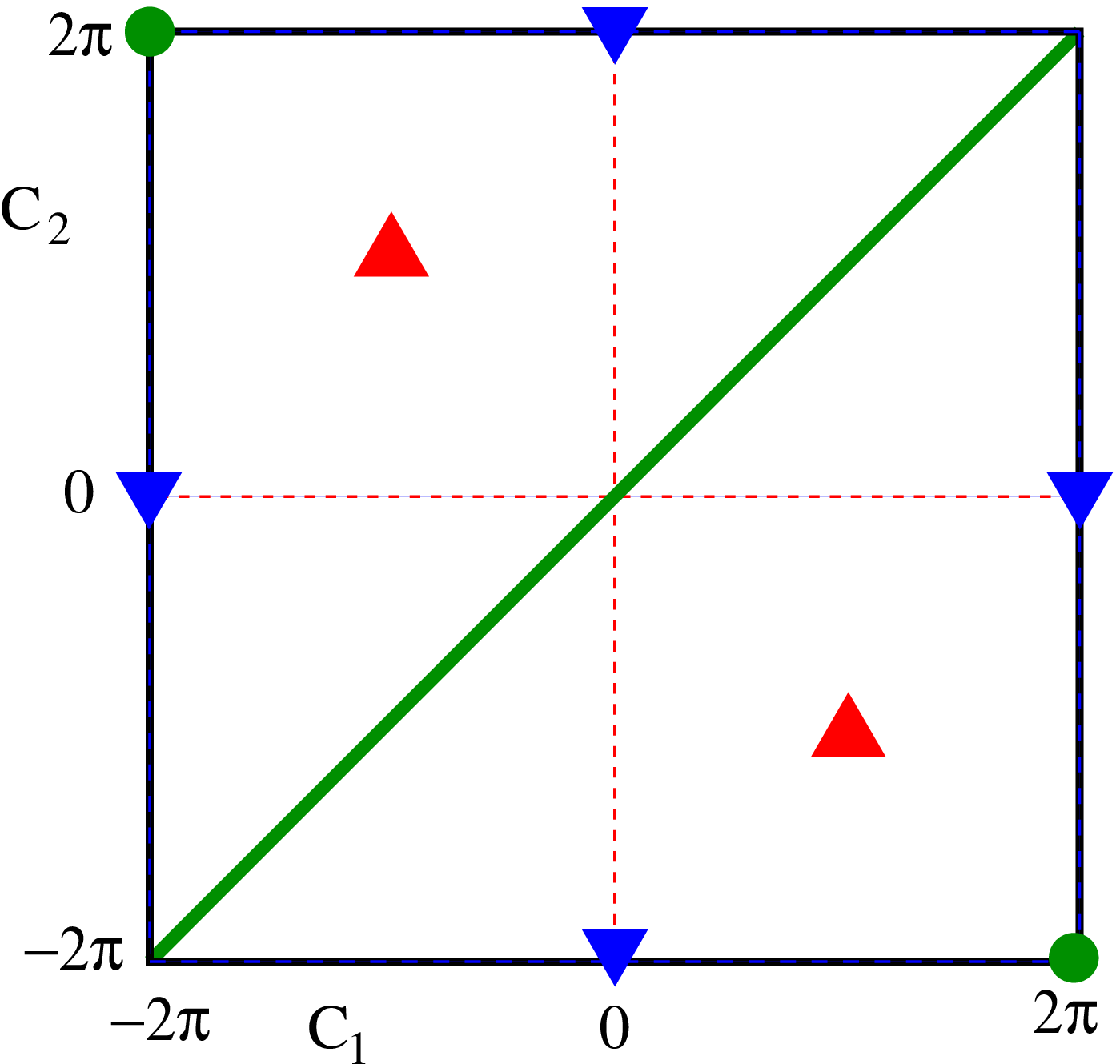}{2truein}

The properties of this potential are similar to those of the orientifold model 
 studied in the previous subsection. Conical minima are again distributed in an intrincate
{\it landscape} with energy spacings of $O(N)$ but reaching energies of $O(N^2)$. 
Conical zeros of the bosonic potentials are given by flat connections of the form
$$
{\vec C}^{(0)}_{1,2} = 4\pi \sum_l {\vec n}_{1,2,l} \,\nu^l
\;.$$
At these points, the fermionic contribution has the value
$$
V_{\rm F} ({\vec C}_1^{(0)}, {\vec C}_2^{(0)}) = 
-N^2 \,V\left({2\pi \over N} \left(\sum {\vec n}_2 - \sum {\vec n}_1 \right) \right)
$$
and, as before, we have negative minima with energies ranging from $O(N)$ to $O(N^2)$
as $\sum {\vec n}_2 - \sum {\vec n}_1$ ranges from $O(1)$ to $O(N)$. 

The global picture is very similar to the case of the orientifold model, including the
rough statistical properties of the potential. The
same analysis used before also shows that the orbifold
potential averages to $O(N)$ energy while the squared potential averages to $O(N^3)$. 

The similarity between the orbifold and the orientifold potential is even quantitative
along the ``antidiagonal" of the moduli space, i.e. the hypersurface defined by
${\vec C}_+ =0$, where  ${\vec C}_{\pm} = \shalf ({\vec C}_1 \pm {\vec C}_2\,)
$ are the eigenvalues of the orbifold ${\bf Z}_2$ action over the moduli coordinates. 
Along this hypersurface of ${\bf Z}_2$-odd 
connections the orbifold potential is identical
to the orientifold one, up to $O(N)$ corrections,
$$
V_{\rm orb} (0, {\vec C}_-) = 2 \,V_{\rm ori} ({\vec C}_-) + O(N)\;. 
$$
Notice that this argument also implies that the origin of moduli space
is unstable along the antidiagonal, in view of the similar behaviour
of the orientifold potential studied in the previous section. However,
there are many other unstable directions, such as ${\vec C}_1=0$,
${\vec C}_2=\nu^1 {\vec c}$, for instance. 

The orbifold potential  also features negative-energy minima with $O(N^2)$
energy that are
 not inherited from down-shifting of the bosonic conical zeros.
In order to exhibit these special minima, 
let us focus on one of the ``outer rims" of the
moduli space, ${\vec C}_2 =0$, ${\vec C}_1 ={\vec C}$, and points of the form
$$
{\vec C}_\epsilon = \left(2\pi \sum_{l=1}^{N-1} \epsilon_l \,\nu^l, 0,0\right)
$$
where $\epsilon_l = \pm 1$, with $O(N/2)$ terms of each sign. The 
 fermionic potential at these points  is
$$
V_{\rm F} ({\vec C}_\epsilon) 
= -2N \sum_i V (\nu^i \cdot {\vec C}_\epsilon\,) \sim -2N \sum_i V(\epsilon_i \pi)
\sim -2N^2\,V(\pi)
\;,$$
whereas the bosonic contribution is at most of $O(N)$.
To see this, notice that 
$$
2\pi \sum_{l=1}^{N-1} \epsilon_l \,\nu^l \cdot (\nu^i - \nu^j) = \pi (\epsilon_i - \epsilon_j)
$$
if $i,j = 1,\dots, N-1$. All these terms are of the form $V(\pm 2\pi)$ or $V(0)$ and
thus vanish. The remaining $N-1$ terms have  argument
$$
2\pi \sum_{l=1}^{N-1} \epsilon_l \,\nu^l \cdot (\nu^i - \nu^N) = \pi \,\epsilon_i
$$
and contribute $\sum_i V(\pm \pi) \sim N \,V(\pi)$ to the bosonic potential. We conclude
that these minima come from the superposition of  $N^2$ inverted maxima of the 
$SU(2)$ potential.

A  peculiar feature of the orbifold 
potential is its exact cancellation along the diagonal ${\vec C}_1 = {\vec C}_2$. This
is the closest we come to the complete cancellation of the potential over the moduli space,
the hallmark of the supersymmetric models. In this case, however, local minima of the potential
lie outside the diagonal on the ``twisted sector" of the moduli space.  
Although in finite volume one cannot strictly talk about spontaneous
breaking of the orbifold ${\bf Z}_2$ symmetry, minimum energy wave functions are localized 
near regions with non-zero values of the ``twisted fields" ${\vec C}_1 - {\vec C}_2$. 
 In  this respect, the status of 
this result is  
similar to that in \refs\rtong, this time in the regime of full spatial compactification.

\newsec{Discussion}

\noindent

In this paper we have examined the finite-volume dynamics of the 
basic examples of orbifold and orientifold 
models that exhibit the property of planar equivalence. We have applied standard 
 methods based on the Born--Oppenheimer approximation to the
 dynamics of asymptotically-free gauge theories on 
small three-tori. In particular, we have analized the large-$N$ properties of
one-loop effective potentials over the moduli space of flat connections. 
This type of analysis leads to the known properties and microscopic determinations of the Witten index in the case of supersymmetric theories. Since planar
 equivalence  consists precisely in a   large-$N$ ``inheritance" 
of certain supersymmetric properties between parent and daughter theories, it
becomes an interesting question whether a ``planar" version of a supersymmetric
index could be defined for these theories.

Examining the graded partition function as a natural candidate for
such a ``planar index", we  have argued that the constraints imposed by
perturbative planar equivalence are not strong enough to justify a useful
notion of planar index, at least in the sense that it should be determined by
$O(N)$ fermionic zero modes 
 for a rank $N$ gauge theory. Instead, we find that effective potentials
over the moduli space of flat connections remain generically too large 
 as $N\rightarrow \infty$ to allow the kind of zero-mode dynamics that is characteristic
of supersymmetric theories. In particular,  potential barriers of $O(N^2)$ height
 remain at special points in the moduli space, resulting in   strong violations  the
much weaker
property of minimal planar equivalence (cancellation of all $O(N^2)$ features).
In general, planar equivalence is only maintained in a statistical sense, at the
level of averages over the moduli space.  

The associated wave functions show 
 exponential localization around local minima, unlike the
constant wave functions that ensure an index determined by fermion zero modes
in the supersymmetric case. 
These large potential barriers also imply that semiclassically calculable
effects on these potentials will be typically suppresed exponentially in the
large-$N$ limit. Identifying a semiclassical tunneling effect of $O(1)$ in
the large-$N$ scaling, but  non-perturbative
suppression in the 't Hooft coupling, would require potential barriers of
order $\Delta V_{\rm eff} \sim 1/LN$, much smaller than those found in
our examples.

The orbifold model admits a very strong projection  under which it behaves
exactly like a supersymmetric theory. This is the diagonal projection, onto
wave functions that are supported only over the diagonal of the moduli space
${\vec C}_1 = {\vec C}_2$. This projection, if exercised over the complete
Hilbert space of the orbifold theory, truncates it to a supersymmetric
$SU(N)$ gauge theory. Notice that the parent theory in this case is $SU(2N)$,
so that we are considering a much stronger projection here. What we find is
that, to one-loop accuracy but exactly over the moduli space of flat connections,
this projection can be done only over the zero-modes with identical result.      

Although this behaviour of the orbifold model is interesting, it is fair
to say that the required diagonal projection is completely {\it ad hoc} and
is not supported by the dynamics of the system, which tends to favour wave
functions supported on the ``twisted" regions of the moduli space. In fact,
our analysis  agrees with previous indications from \refs\rtong, suggesting that
the  ${\bf Z}_2$ symmetry might be spontaneusly broken in infinite volume, a
situation that would rule out the planar equivalence realized in orbifold-type
models.  

It would be interesting to complement our analysis with the study of more
general boundary conditions, allowing the presence of magnetic fluxes through
the torus. In the case of the supersymmetric theories, 
this introduces interesting
refinements of  the index that probe non-trivial properties such as confinement
and chiral symmetry breaking. In the context of the
 models considered in this paper, the study of such twisted sectors could reveal
a better large-$N$ vacuum behaviour in the orbifold case, and perhaps a hint
of the fermion condensates that should develop in the infinite-volume theory
in the case of the orientifold model. Work in this direction is in progress
(c.f. \refs\rcarlosp).   

\vskip 1cm

{\bf Acknowledgements}

\noindent

We would like to thank A. Armoni and C. G\'omez for enlightening discussions.
The work of C.H. was partially supported by an FPU grant from MEC-Spain.
 The work of J.L.F.B. was partially supported by MCyT
 and FEDER under grant
BFM2002-03881 and
 the European RTN network 
 MRTN-CT-2004-005104.

\listrefs

\bye